\documentclass{elsart}

\usepackage{amssymb}
\usepackage{graphicx}
\begin{document}

\begin{frontmatter}



\title{A Multiwire Proportional Chamber for Precision Studies of
  Neutron $\beta$ Decay Angular Correlations}

\author{T.~M.~Ito\corauthref{cor1}\thanksref{label1}}
\corauth[cor1]{Corresponding author}
\ead{ito@lanl.gov}
\thanks[label1]{Present address: Los Alamos National Laboratory, Los
  Alamos, NM 87545} 
\author{R.~Carr},
\author{B.~W.~Filippone},
\author{J.~W.~Martin\thanksref{label2}},
\thanks[label2]{Present address: Department of Physics, The University
  of Winnipeg, Winnipeg, Manitoba, R3B 2E9} 
\author{B.~Plaster},
\author{G.~Rybka\thanksref{label3}},
\thanks[label3]{Present address: Massachusetts Institute of
  Technology, Cambridge, MA} 
\author{J.~Yuan\thanksref{label4}}
\thanks[label4]{Present address: Harvard University, Cambridge, MA} 
\address{W.~K.~Kellogg Radiation Laboratory, California Institute of
  Technology, Pasadena, CA 91125}

\begin{abstract}
A new multiwire proportional chamber was designed and constructed for
precision studies of neutron $\beta$ decay angular correlations. Its
design has several novel features, including the use of low pressure
neopentane as the MWPC gas and an entrance window made of thin Mylar
sheet reinforced with Kevlar fibers. In the initial off-line
performance tests, the gas gain of neopentane and the position
resolution were studied.

\end{abstract}

\end{frontmatter}

\section{Introduction}
\label{sec:Intro}
Precision studies of neutron $\beta$ decay offer an excellent means to
test the foundation of the Standard Model of electroweak interactions
and to search for what may lie beyond it. In particular, the angular
correlation between the neutron spin and the electron momentum
(characterized by the coefficient $A$) in polarized neutron $\beta$
decay provides important input for testing the unitarity of the CKM
matrix~\cite{ABE04}. Recent technological advancements, including the
realization of superthermal ultracold neutron sources~\cite{MOR02,SAU04},
have made experiments with a precession of $\delta A/A \sim 10^{-3}$ within
reach. Consequently, it is imperative to control possible systematic
effects at the 0.1\% level.

A typical experimental arrangement for $A$ coefficient measurements
involves measuring the forward-backward asymmetry of electron emission
with respect to the neutron spin in polarized neutron $\beta$
decay. Possible sources of systematic effects include polarization
determination, background, and detector effects such as electron
backscattering. With regard to the detector effects, most of the
previous experiments used plastic scintillation counters as the
detector for the decay electron. However, due to the small end point
energy of the electron spectrum ($E_0=782$~keV), a significant fraction
($\sim 10\%$) of electrons from neutron $\beta$ decay directed to one
detector can backscatter from the surface of the detector and are detected
by the other detector. A non-negligible fraction of the backscattered
electrons
leave undetectably small energy deposition in the first detector,
hence introducing an error in the asymmetry determination. These
electrons are called missed backscattered electrons.

One way to reduce the fraction of the missed backscattered electrons,
and hence the effect of such events, is to use a detector that is
sensitive to smaller energy deposition and can therefore have a lower
energy threshold. In general, gas proportional counters
are sensitive to a smaller energy deposition than plastic
scintillation counters are. For a gas counter, an energy deposition of
$\sim 100$~eV is necessary to create one primary electron-ion pair,
which generates on average four secondary electron-ion pairs. With a
moderate gas gain of 10$^4$ and an amplifier gain of 10~V/pC, one
obtains an output pulse of $\sim 50$~mV. Therefore, an energy
deposition of 100~eV can be relatively easily detected. On the other
hand, for a plastic scintillation detector, an energy deposition of
100~eV is necessary to create one photon inside the scintillator. If a
typical value of 15\% for the photocathode quantum efficiency of the
photomultiplier tube (PMT) and a rather generous value of 25\% for the
light collection efficiency are assumed, we obtain a 2.5~keV energy
deposition for a single photoelectron event.

A detector system comprised of a thin low pressure multiwire
proportional chamber (MWPC) placed in front of a plastic scintillator
serves as an ideal $\beta$ detector: the thin MWPC provides
sensitivity to backscattered events with small energy deposition while
the plastic scintillator provides the total energy information as well
as the timing information, which is of vital importance in analyzing a
class of events in which both detectors record a hit as a result of
backscattering. In addition, the position information from the MWPC
can be used to apply a fiducial volume selection, which is of vital
importance in experiments that use ultracold neutrons (UCN).  Also, by
requiring a coincidence between the MWPC and the scintillation counter
for the event trigger, the sensitivity to $\gamma$ rays, one of the
major background sources in previous experiments, can be reduced
because of the relatively low sensitivity of MWPCs to $\gamma$ rays.

The UCNA experiment~\cite{UCNA}, currently being commissioned at Los
Alamos National Laboratory, aims at a 0.2\% measurement of the $A$
coefficient using UCN, and uses such a combination as the $\beta$
detector.

In this paper, we describe the design and construction of the MWPC
developed for the UCNA experiment, and report the results from the
initial offline performance tests. The design and construction of the
plastic scintillation detector, together with the performance of the
MWPC as installed in the UCNA spectrometer will be reported in a later
publication~\cite{PLA06}.

\section{Design and construction}
\subsection {General considerations and design overview}
\begin{figure}
\centering
\includegraphics[bb= 100 100  500 700, width=10cm,angle=90]{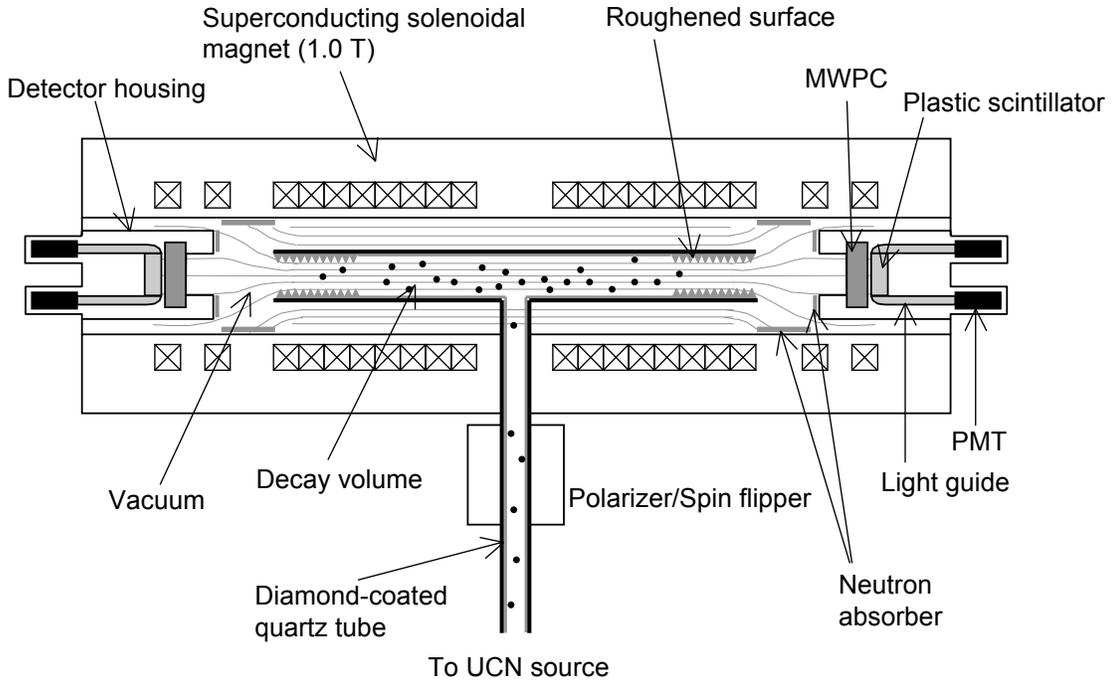}
\caption{Schematic of the UCNA Experiment. 
}
\label{fig:ucna_schematic} 
\end{figure}

In the UCNA experiment, UCN are produced by the LANSCE solid
deuterium UCN source, sent through a polarizer/spin flipper, and then
introduced into a decay volume. The wall of the decay volume is a
3~m-long diamond coated quartz cylinder 10~cm in diameter.  The decay
volume is in the warm bore of a superconducting solenoidal magnet
(SCS), which provides a holding field of 1~T. The decay electrons
spiral along the magnetic field lines towards one of the detectors,
and then enter the field expansion region, where the magnetic field is
reduced to 0.6~T. As an electron enters the field expansion region,
the energy associated with the angular motion of the electron is
transfered to longitudinal motion in order to conserve angular
momentum as the diameter of the spiral increases due to the reduced
field. This reduces the incident angle of the electron onto the
detector surface (reverse of the magnetic mirror effect) and
suppresses backscattering. The detectors are placed in a region where
the expanded field is uniform. The schematic of the UCNA experiment is
shown in Fig.~\ref{fig:ucna_schematic}.

The function of the UCNA MWPC is two-fold: 1) to minimize the missed
backscattered events, and 2) to provide position information. The
latter is necessary because, unlike the previous $A$ measurements that
used a cold neutron beam, in the UCNA experiment the UCN fill the
decay volume. Therefore it is important to place a fiducial volume
cut to reject events in which the $\beta$ decay occurred near the wall
of the decay tube where it is likely for the electron to hit the wall.

A general consideration for minimizing missed backscattered events
results in the following requirements for the design of the MWPC, in
particular for the windows and the gas.
\begin{itemize}
\item The entrance and exit windows have to be thin.
\item The gas has to be at low pressure, and has to consist of relatively
  heavy molecules made of low $Z$ (=atomic number) atoms.
\end{itemize}
The requirements for the windows is obvious as any event in which the
electron gets backscattered from the entrance window without getting
into the gas volume will become a missed backscattered event. The gas
pressure has to be low in order to make the entrance window thin,
since it separates the MWPC gas volume from the spectrometer vacuum.
Also, in order to minimize the backscattering, the gas should only
contain atoms with low $Z$ (=atomic number), since the total
backscattering fraction is larger for a larger $Z$~\cite{TAB71}.  At
the same time, in order to have a high enough detection efficiency,
the MWPC gas has to have a reasonable electron density. This calls for a
gas that consists of relatively heavy molecules made of light atoms.

Furthermore, there are two requirements for the UCNA MWPC design that
are specific to the UCNA experiment. These are requirements for the
position resolution and the actual geometrical size of the MWPC.
\begin{itemize}
\item The maximum diameter of the spiral of the electron trajectory in
the field expansion region is 6.6 mm. In order to perform a fiducial
volume selection with a minimum loss of statistics, a position
resolution of $\sim 2$~mm is required.
\item Because of the field expansion factor of 0.6 (0.6 T/1.0 T), the
10~cm diameter cross section of the decay volume maps to a circle with
$\sim 14$~cm diameter in the field expansion region. On the other
hand, the diameter of the warm bore the SCS magnet is
25.4~cm. Therefore, the UCNA MWPC has to have an effective area larger
than 14~cm in diameter and has to fit in the warm bore with a diameter
of 25.4~cm.
\end{itemize}

There is an additional consideration associated with the exit
window. In the UCNA experiment, the exit window separates the MWPC gas
volume from the volume in which the plastic scintillation detector is
placed. The exit window can be made thin by filling the volume in
which the plastic scintillator is placed with a gas that is of the
same pressure as the MWPC gas.

Below, we will discuss each of these aspects of the design in more
detail.

\subsection{Mechanical construction}
\label{sec:mechanical}
\begin{figure}
\centering
\includegraphics[bb= 0 150  450 750,width=10cm,angle=90]{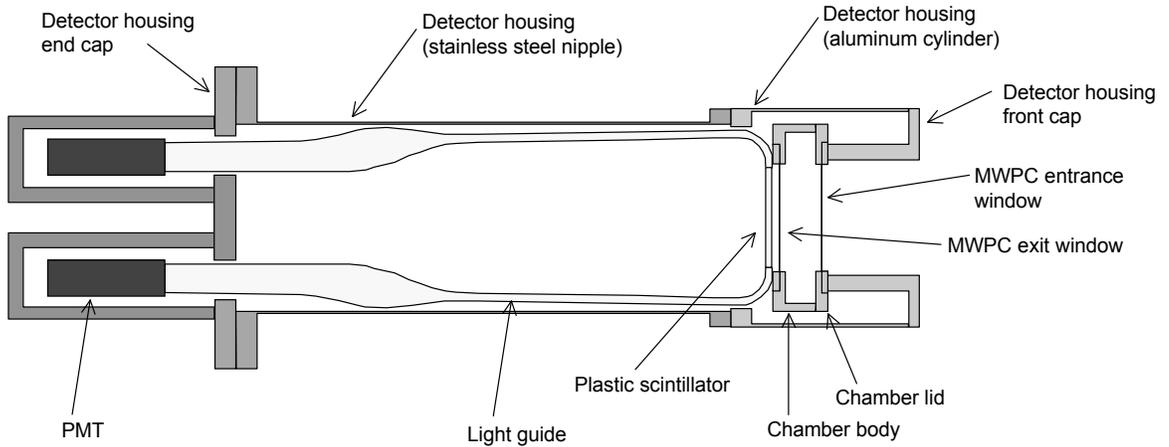}
\caption{Cross section of the detector assembly inserted into the
vacuum inside the SCS warm bore. }
\label{fig:mechanical} 
\end{figure}

\begin{figure}
\centering
\includegraphics[bb= 0 0  590 840,width=10cm,angle=90]{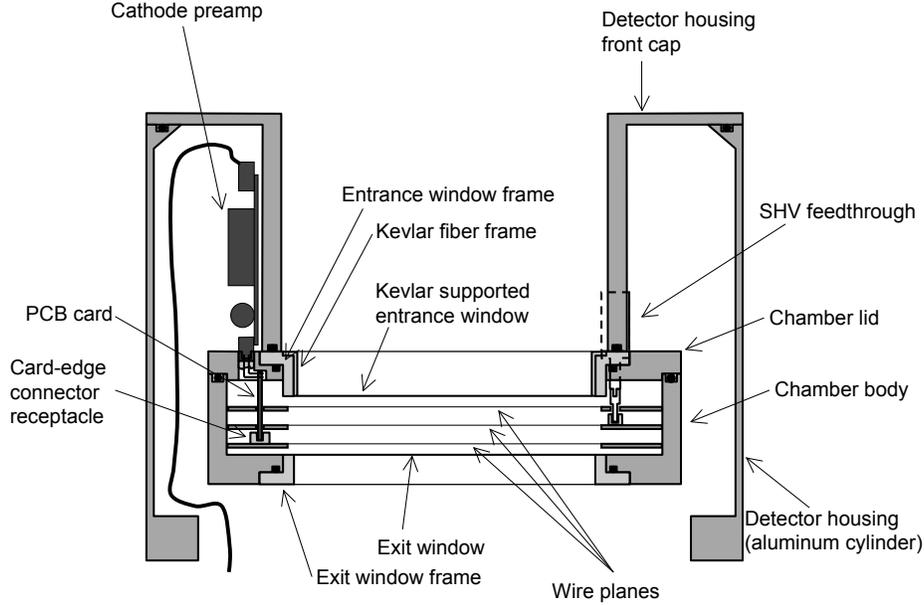}
\caption{Cross section of the MWPC. When the chamber lid is pushed
  against the chamber body, the electrical connection between the wire
  planes and the feedthrough connectors on the chamber lid are made,
  as well as the o-ring seal between the chamber body and the chamber
  lid. For the cathode signals, the connection between the chamber lid
  and the wire planes are provided by a PCB card that slides into a
  card-edge connector receptacle mounted on the FR4 glass epoxy frame
  for the wire plane. The contact pads on the PCB cards are soldered
  to pins on a flat ribbon cable connector receptacle that is glued
  into an opening on the chamber lid serving as a feedthrough
  connector. For the anode, the connection between the wire plane and
  the chamber lid is provided by a metal pin mounted on the SHV
  feedthrough connector sliding into a receptacle mounted on the FR4
  glass epoxy frame of the anode wire plane. See also
  Fig.~\ref{fig:mwpc_pic2}.}
\label{fig:mechanical_details} 
\end{figure}

\begin{figure}
\centering
\includegraphics[width=10cm]{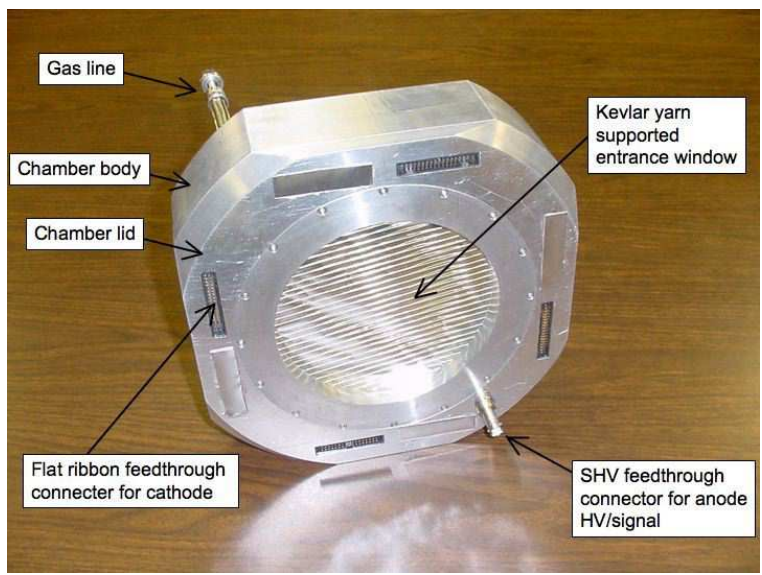}
\caption{A photograph of the MWPC. The chamber lid is mounted on the
  chamber body to enclose the MWPC gas volume. The window and the
  Kevlar support are both installed on the chamber lid (see
  Section~\ref{sec:windows}).}
\label{fig:mwpc_pic1} 
\end{figure}

\begin{figure}
\centering
\includegraphics[width=9cm]{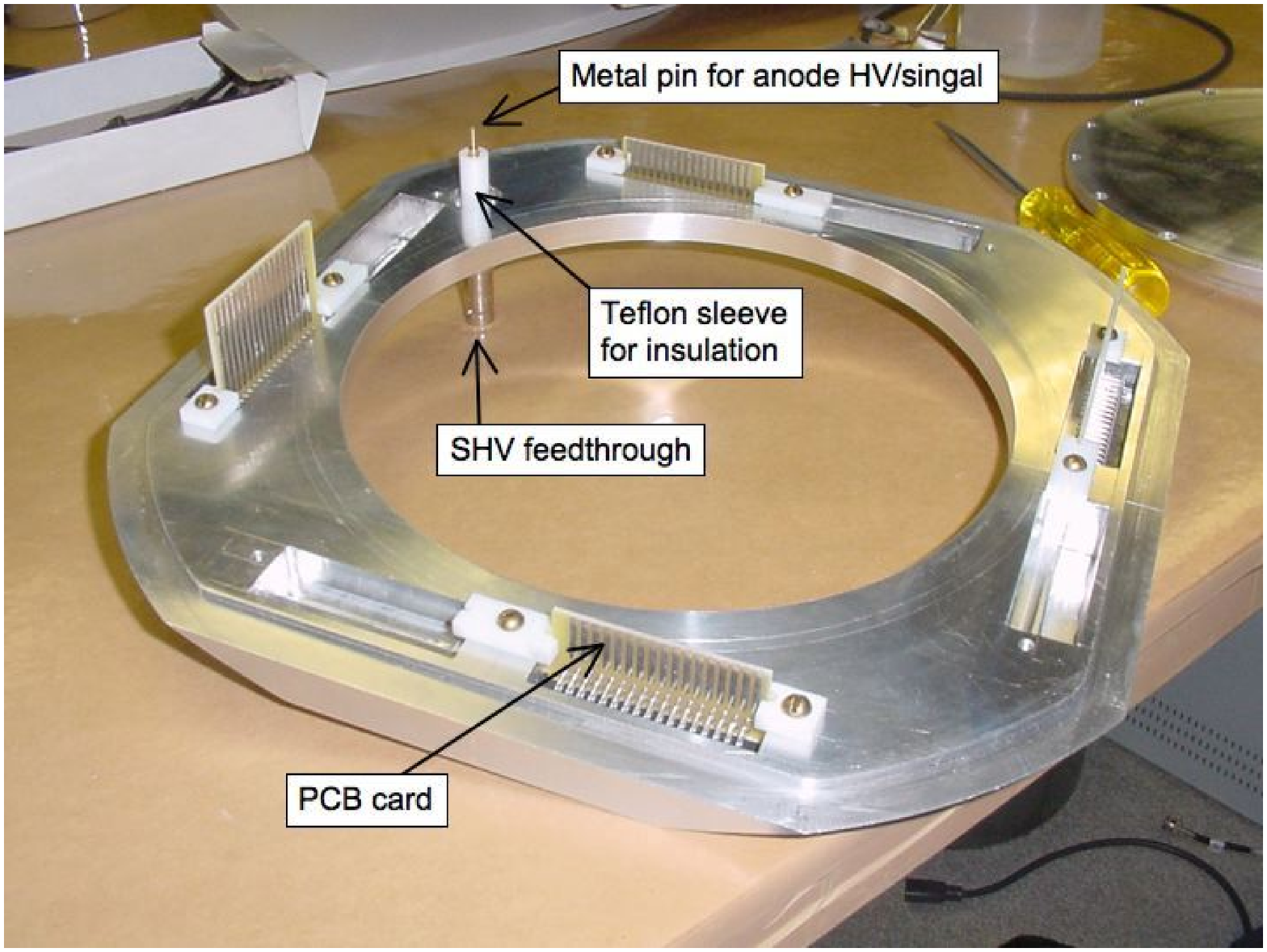}
\includegraphics[width=9cm]{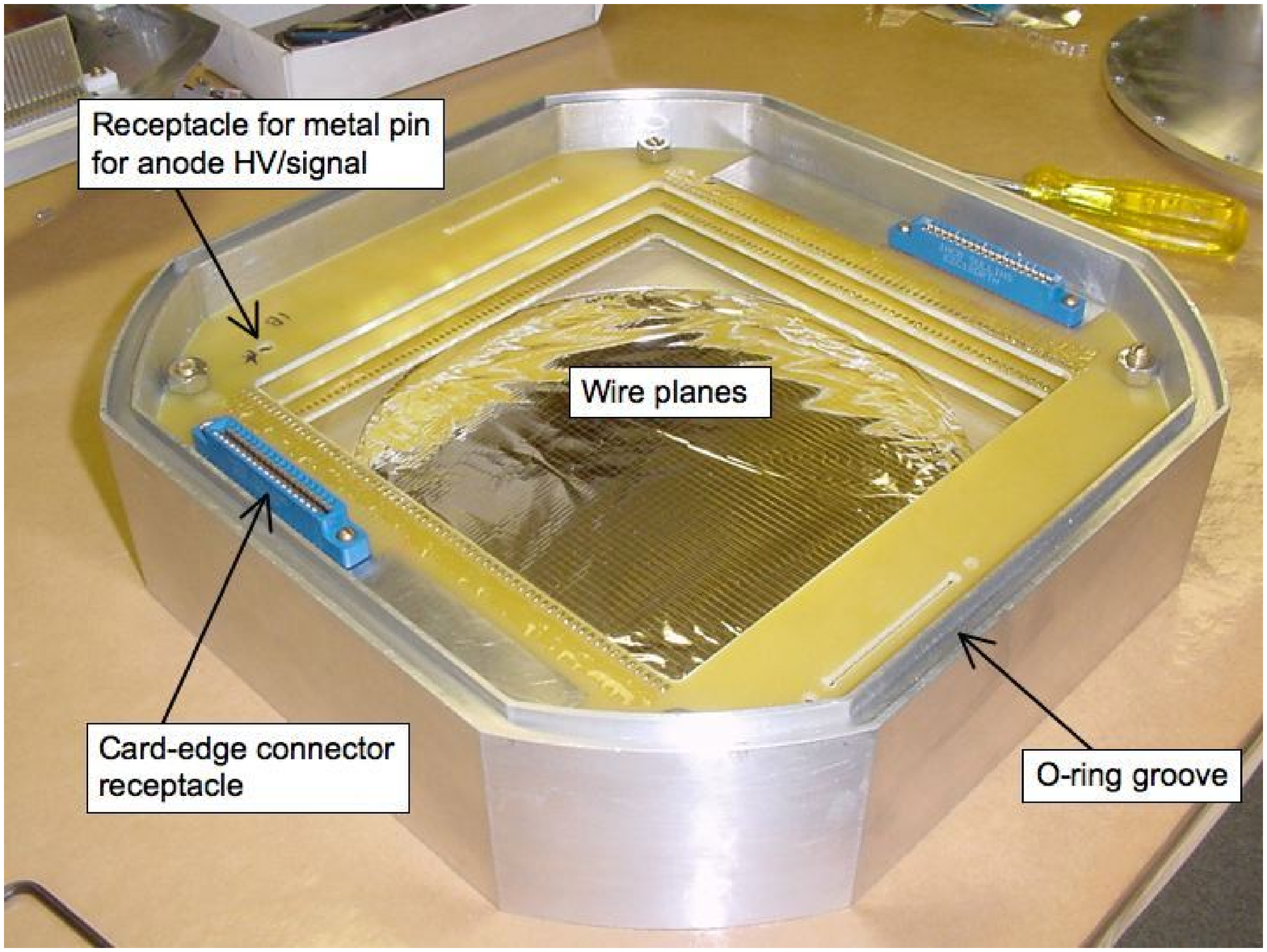}
\caption{Photographs of the chamber lid (top) and the chamber body
    (bottom). The PCB cards on the chamber lid slide into the
    card-edge connector receptacle on the cathode wire plane frames,
    providing the electrical connection for cathode signals between
    the wire planes and the feedthrough connectors on the chamber
    lid. (The longer PCB cards are for the bottom cathode plane and
    the short ones are for the top cathode plane.) Also the metal pin
    mounted on the SHV feedthrough on the chamber lid slides into a
    receptacle on the anode plane frame, providing the electrical
    connection for the anode HV and signals between the wire plane and
    the chamber lid.}
\label{fig:mwpc_pic2} 
\end{figure}

Figure~\ref{fig:mechanical} shows the cross section of the detector
assembly, which in normal operation is inserted into the vacuum inside
the SCS warm bore. Figure~\ref{fig:mechanical_details} shows the cross
section of the MWPC and Figs.~\ref{fig:mwpc_pic1} and
\ref{fig:mwpc_pic2} are photographs of the MWPC.  The chamber body and
the chamber lid, both made of aluminum, constitute the body of the
MWPC itself and enclose the MWPC gas, which is 100 torr neopentane
(see section~\ref{sec:gas}). The MWPC, the plastic scintillator, the
light guide system, and the photomultipliers are housed in the
detector housing. The detector housing consists of a about 25 in. long
stainless steel nipple with an outer diamter of 10.5 in., an aluminum
cylinder with an outer diameter of about 12 in., an end cap and a
front cap made of aluminum. The stainless steel nipple and the
aluminum cylinder are connected via 12-in. conflat flanges. The
aluminum cylinder is welded onto a 12-in. flange made of explosion
bonded aluminum-stainless steel bi-metal.

The detector housing is filled with 100 torr nitrogen.  This is to
minimize the pressure difference across the MWPC back window so that
the back window can be thin.  The space enclosed by the aluminum
cylinder, the front cap, and the chamber lid houses the preamplifier
for the MWPC readout and the HV systems for the MWPC operation.  The
difficulties associated with operating the HV system in a 100 torr
nitrogen environment will be discussed in Section~\ref{sec:hv}.

The chamber body, the chamber lid, the front cap are designed so that
when the front cap is mounted on the aluminum cylinder it pushes the
chamber lid making the o-ring seal between the chamber body and the
chamber lid, the o-ring seal between the chamber lid and the front
cap, and the o-ring seal between the front cap and the aluminum
cylinder. Also it makes the electrical connection between the
connectors on the chamber lid and the wire planes when the front cap
is mounted on the aluminum cylinder pushing the chamber lid on the
chamber body (See Figs.~\ref{fig:mechanical_details} and
\ref{fig:mwpc_pic2}).

\begin{figure}
\centering
\includegraphics[width=10cm]{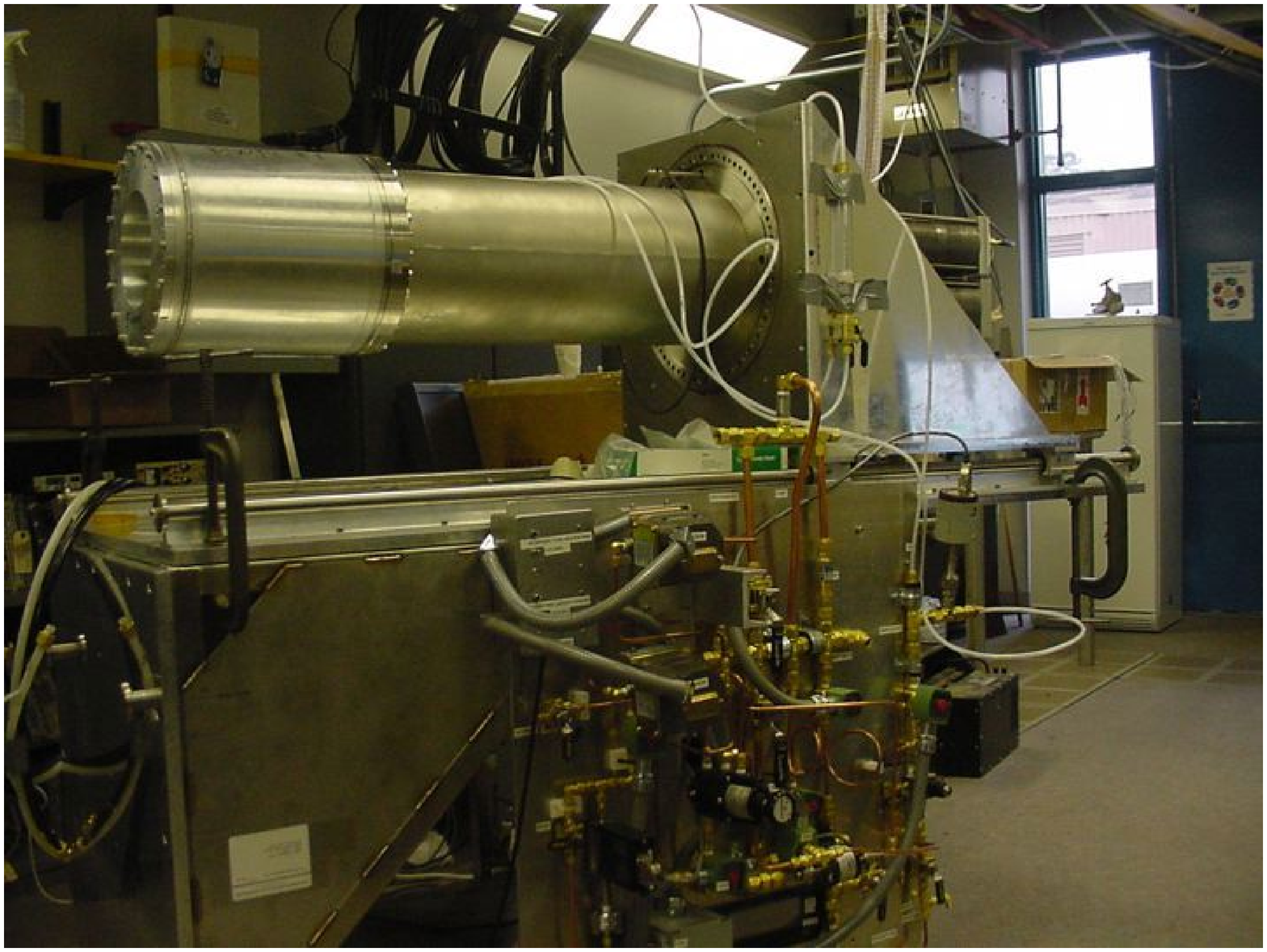}
\caption{A photograph of the detector assembly and the gas panel.}
\label{fig:detector_assembly_pic} 
\end{figure}

Shown in Fig.~\ref{fig:detector_assembly_pic} is a photograph of the detector
assembly and the gas panel. 


\subsection{Wire chamber gas}
\label{sec:gas}
The requirements discussed earlier that the wire chamber gas consist
of relatively heavy molecules made of low $Z$ atoms and practical
requirements (availability, affordability, etc) limited our choices
for the gas almost uniquely to heavy hydrocarbon gases.

We chose neopentane (C$_{5}$H$_{12}$, also known as
2,2-Dimethylpropane) because of the high vapor pressure at room
temperature (1125~torr). Typically we operated the wire chamber at 100
torr, which is safely below the vapor pressure. Also, as discussed
below, we were able to develop a thin window that can safely withstand
a 100 torr pressure difference. Neopentane at 100~torr has a electron
density of $1.5\times 10^{20}$/cm$^3$. For comparison, P10 gas at 1
atm has an electron density of $4.6\times 10^{20}$/cm$^3$. Thus we
expect a reasonably similar detection efficiency for 100 torr
neopentane compared to 1 atm P10 gas, provided that we get a
reasonably good gas gain and charge collection.

We note that n-pentane (normal pentane, C$_{5}$H$_{12}$) has already
been used as a proportional chamber gas. Ref~\cite{LAZ98} discusses
the use of gas mixtures of n-pentane-CO$_2$, n-pentane-CO$_2$-CH$_4$,
and n-pentane-CO$_2$-CF$_4$, while Ref.~\cite{BRE79} discusses the use
of various hydrocarbon gases such s isobutane, ethylene, n-pentane,
and n-heptane at very low pressures ($0.3-10$ torr). Although
n-pentane is more cost effective than neopentane, it is not a
practical choice for the following reasons:
\begin{itemize} 
\item The high reactivity, which would considerably limit the material
allowed for the construction of the MWPC.
\item The low vapor pressure (424~torr at 20$^{\circ}$C), which would
make the operation of the gas panel more demanding to avoid accidental
liquefaction of the gas.
\end{itemize}

We will discuss the measured properties of neopentane as a wire
chamber gas in Section~\ref{sec:gasproperties}

\subsection{Wire plane construction}
\label{sec:wireplanes}
Heavy hydrocarbon gas such as isobutane is commonly used as a quencher
gas, not as the main component of the wire chamber gas. In order to
obtain a high enough gain from neopentane, which is also a heavy
hydrocarbon gas, a wire with a small diameter has to be used for the
anode wire plane. We chose to use gold-plated tungsten wires with a
diameter of 10~$\mu$m.  64 anode wires are strung with a 2.54~mm spacing,
giving an effective area of 163~mm $\times$ 163~mm, and are soldered
onto a 0.092-in. thick FR4 glass epoxy frame. A tension of 9~gf, which
is about 80\% of the breaking point of the wire, was applied to each
wire. All the anode wires are connected together by conductor tracks
etched onto the FR4 glass epoxy plate, which are then connected to the
HV resistor located outside the MWPC through an SHV feedthrough
connector mounted on the chamber lid (see Sections~\ref{sec:mechanical}
and \ref{sec:hv} and Fig.~\ref{fig:mechanical_details}).

In order to minimize electron backscattering, we chose to use wires also
for the cathode planes, instead of more conventionally used thin
strips of conductor etched on a substrate such as Mylar
sheet~\cite{MYL}. The entrance and exit windows cannot be used as the
cathode planes because the thin windows bow out by up to 1~cm (as
measured at the center) due to the pressure difference across the
windows. Also, the use of wires instead of conductor strips etched on
a thin substrate facilitated the pump-out operation necessary prior to
filling the MWPC with 100 torr neopentane.

Gold-plated aluminum wires 50~$\mu$m in diameter were used for the
cathode planes. Each cathode plane has 64 wires, strung with a 2.54~mm
spacing with a 50~gf tension and soldered onto a 0.092-in. thick FR4
glass epoxy frame. Every four neighboring cathode wires are connected
to one conductor strip etched on the FR4 glass epoxy plate, which then
was connected to a card edge connector receptacle mounted on the FR4
epoxy plate. (There are 16 strips on each cathode plane.)  The card
edge connectors then provide a connection to the outside of the MWPC
through feedthrough connectors on the chamber lid, on which Multi
Channel Systems CPA16 amplifier carrier boards are mounted~\cite{MCS}
(See Sections~\ref{sec:mechanical} and \ref{sec:readout}).

Wire planes were constructed using a wire-winding machine built
according to Ref.~\cite{KUZ87}. The wire-winding machine was
originally built for time projection chambers for the DRIFT dark
matter search experiment~\cite{SNO00}.  We modified it replacing the
weight that provides a constant tension to the wire with a hysteresis
clutch/brake~\cite{Ogura} which provides a constant torque, for a
simpler operation. The wire was wound onto a frame made of aluminum,
and then transfered onto the G10 frame.

The distance between the anode plane and a cathode plane was chosen to be
10~mm, about 4 times the anode wire spacing~\cite{SAU77}.  The two
cathode planes are mounted in such a way that the wires on one plane
run perpendicular to the wires on the other plane, thus providing
positional information in both directions that are perpendicular to
the axis of the SCS spectrometer (see Fig.~\ref{fig:readout}).

\subsection{Bias High Voltage Systems}
\label{sec:hv}
In order to collect the charge deposited between the windows and the
cathode planes, the cathode planes are designed so that they can be
held at a positive potential (up to $\sim$ 400~V) with respect to the
windows which are grounded. The anode plane then is held at a positive
potential ($\sim$ 2700 V) with respect to the cathode planes.

The anode bias high voltage is supplied through a 25 M$\Omega$
resistor (See Fig.~\ref{fig:readout}). A 6.8 nF HV capacitor was
inserted between the ground and the HV and served as a filter. The 25
M$\Omega$ resistor, the 6.8 nF capacitor, and the 3.3 nF decoupling
capacitor were all enclosed in a box (``HV box'') that was filled with
Araldite 2011 epoxy~\cite{ARA} to prevent discharge due to the low
pressure (100 torr nitrogen) environment.

The high voltage to the HV box was provided through an SHV cable that
must be operated in the 100 torr nitrogen atmosphere. We found that
most of the commercially available SHV cables failed in this low
pressure environment. Detailed tests revealed that for some cables
this failure was due to the SHV connectors themselves (SHV connectors
from some manufactures always failed), and for others this was due to
the way cables were assembled. We used SHV cables that were obtained
from Dielectric Sciences~\cite{DIE}.

\subsection{Readout method and electronics}
\label{sec:readout}
\begin{figure}
\centering
\includegraphics[bb= 0 0  590 840,width=10cm,angle=90]{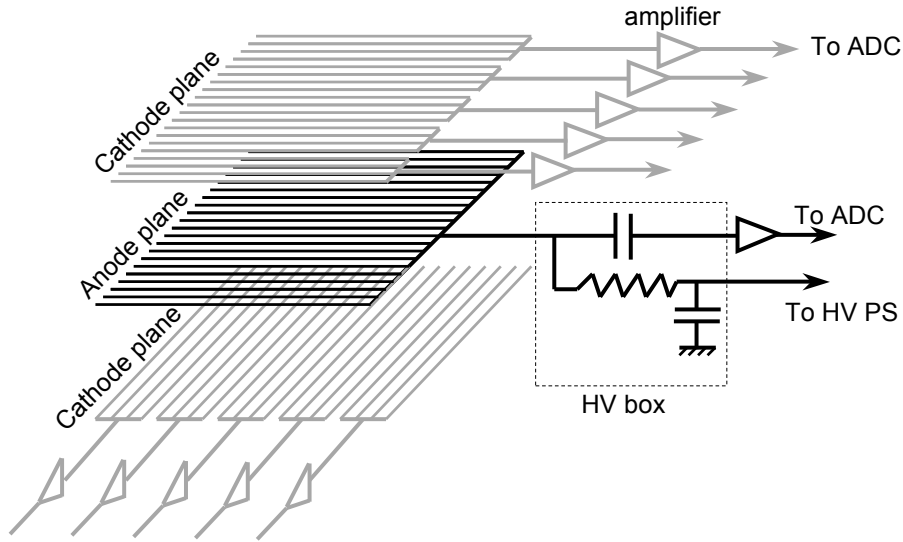}
\caption{Schematic showing the readout method for the anode and
  cathode wire planes. The anode wires are all ganged together,
  providing the information on the total energy deposited to the
  MWPC. Every four cathode wires are ganged together, with the center
  of gravity of the cathode signals providing the location of the
  hit.}
\label{fig:readout} 
\end{figure}

The basic readout method is shown in Fig.~\ref{fig:readout}.  As
discussed in section~\ref{sec:wireplanes} the signals from all the
anode wires are summed on the anode wire plane, providing the
information on the total energy deposited in the MWPC. The signals
from every four neighboring cathode wires are summed, giving 16
channels to read out for each cathode plane. For the given
cathode-anode plane distance (10~mm), the distribution of the induced
charge on the cathode planes has a width of about 2~cm (FWHM), thus
three to four cathode strips (each strip is a collection of four
neighboring cathode wires) receive a detectable signal for each hit on
the MWPC. The center of gravity of the signals of each cathode plane
provides the coordinate of the location of the avalanche in the
direction perpendicular to the cathode wire. Combining the information
from both cathode planes determines the location of the avalanche.

The summed anode signal is extracted to the outside of the MWPC
through an SHV feedthrough connector and fed to a Multi Channels
System PA3300 amplifier module through a decoupling capacitor of
3.3~nF (See Fig.~\ref{fig:readout}). The PA3300 module contains a
charge-sensitive preamplifier and a shaping amplifier. The gain and
the shaping time are chosen to be 2~V/pC and 0.25~$\mu$s,
respectively.

The signal from each of the 16 cathode strips is also extracted to the
outside of the MWPC and is fed to a Multi Channel Systems PA3300
amplifier. For the cathode, the 16 PA3300 amplifiers are mounted on a
Multi Channel Systems CPA16 carrier board, which provides a decoupling
capacitor (rated up to 500~V) for each channel as well as power for the
PA3300 amplifiers. For the cathode, the gain and shaping time for the
PA3300 amplifiers are chosen to be 25~V/pC and 0.25~$\mu$s,
respectively.

The amplified anode and cathode signals are digitized by a CAEN peak
sensing multi-event ADC V785~\cite{CAEN}.

\subsection{Thin entrance and exit windows}
\label{sec:windows}
\begin{figure}
\centering
\includegraphics[bb= 0 0  590 840,width=10cm,angle=90]{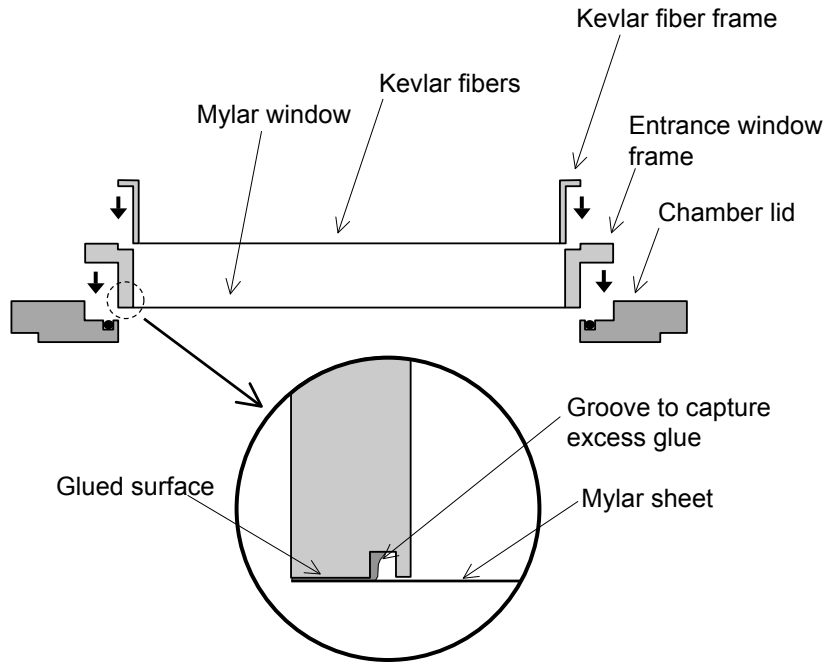}
\caption{Schematic showing the structure of the entrance window and
  the Kevlar support.}
\label{fig:window} 
\end{figure}

The entrance window is 15~cm in diameter and is made of 6~$\mu$m thick
aluminized Mylar sheet, reinforced by 200 denier Kevlar
fiber~\cite{KEV}. The window is glued onto the window frame with
Araldite 2011 epoxy, and is then inserted into the opening on the
chamber lid. An o-ring provides a seal between the window frame and the
chamber lid. The Kevlar fiber is strung with a 5~mm spacing, and is
epoxied onto a separate aluminum frame which is inserted into the opening of
the window frame (See Fig.~\ref{fig:window}). 

During the developmental tests, we found that it was very important not
to let the excess epoxy flow out onto the surface of the Mylar sheet
when the window frame is glued on to a Mylar sheet. The epoxy hardened
on the surface of the Mylar sheet causes a tear to develop when the
window is subject to a pressure difference and needs to bow in or
out. Machining a groove on the window frame to capture the excess
epoxy (see Fig.~\ref{fig:window}) substantially improved the longevity
of the windows.

Thin Kevlar reinforced Mylar windows are discussed in
Ref.~\cite{ATE93}. However, our method differs from that in
Ref.~\cite{ATE93} as follows. In Ref.~\cite{ATE93}, the window is of a
cylindrical shape, and therefore the best performance was obtained
when the Kevlar fibers were glued on the surface of Mylar sheet
preformed to a cylindrical shape. In our case, the window is glued
onto a flat circular frame, and yields to some radius when a pressure
difference is applied across the window. If the Kevlar fibers are
directly glued on to the surface of the Mylar sheet, it tends to rip
the Mylar sheet off as the window bows out under a pressure
difference. The best performance was obtained when the Kevlar fibers
are only fixed on the frame, allowing them to move against the surface
of the Mylar sheet when it bows out under a pressure difference.

The entrance window withstands a pressure difference of up to
200~torr. We chose this configuration as a result of a compromise
between the strength (and the leak rate) and the amount of
material. For example, 3~$\mu$m Mylar sheet could not be used because
of the high leak rate due to pin holes on the sheet.

The exit window is 15~cm in diameter and is made of 6~$\mu$m thick
aluminized Mylar sheet. Since there is only a small pressure
difference ($\sim 5$~torr), there is no need for Kevlar reinforcement.

\subsection{Gas handling system}
The function of the gas handling system is as follows.
\begin{itemize}
\item In normal operation, it maintains the MWPC gas pressure at
  100~torr within 1\%, while flowing gas into the MWPC at a rate of
  $\sim 10-15$~cc/min. It also maintains the pressure in the nitrogen volume to
  be $\sim 5$~torr below the MWPC gas pressure.
\item During the initial pump-out, it evaculates the MWPC and the
  nitrogen volume while keeping the pressure difference across the
  front and back window within the limit given for each window.
\item It protects the MWPC windows and the wire planes in case of
  accidents, including a loss of the SCS vacuum and a power failure.
\end{itemize}

\begin{figure}
\centering
\includegraphics[bb=0 0 600 800, width=10cm,angle=90]{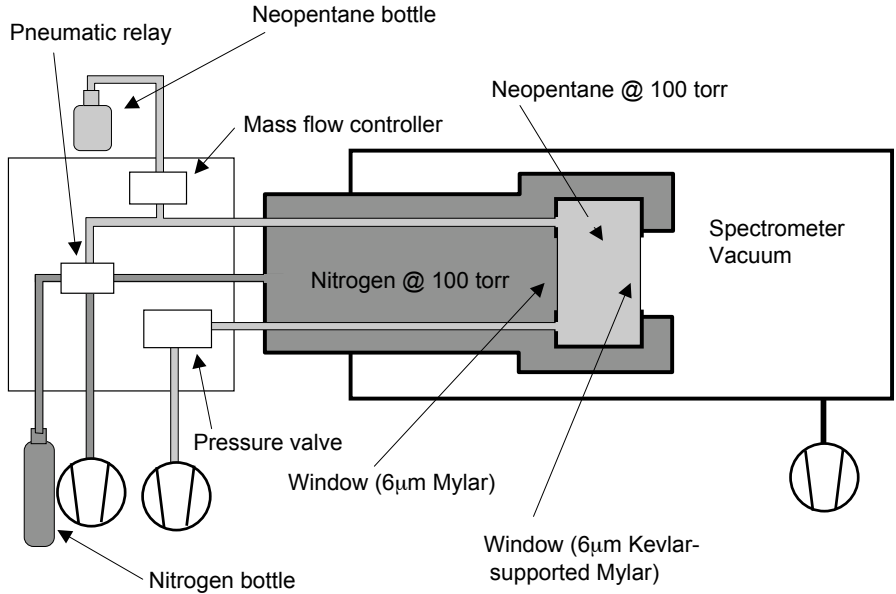}
\caption{Schematic of the gas flow}
\label{fig:gaspanel1} 
\end{figure}

Figure~\ref{fig:gaspanel1} shows how the constant gas flow and the
constant pressure in the MWPC volume and the constant gas pressure in
the nitrogen volume are maintained during normal operation (the
first item listed above). The mass flow controller (MKS type
1179A)~\cite{MKS} ensures that a constant mass flow of gas enters the
MWPC gas volume, while the pressure valve (MKS type 248A) throttles
its opening to control the flow of gas that leaves the MWPC gas volume
so that the pressure is maintained at the target pressure
(100~torr). The pressure of the gas in the MWPC gas volume is measured
by a capacitance manometer (MKS Baratron 626~\cite{BAR}), and the
pressure information is used by a controller (MKS type 146C) that
controls the pressure valve.

The nitrogen pressure was controlled by a pneumatic relay (Fairchild
Model 14212EJ). It uses the control pressure (in this case the MWPC
gas pressure) to control the output pressure (in this case the
nitrogen pressure). There is a constant flow of nitrogen gas through
the input. The excess gas is directed to a vacuum pump.

With this system, we are able to control the MWPC gas to maintain its
pressure at 100~torr within 0.5\%. The nitrogen pressure in the
nitrogen volume is typically maintained at 5~torr below the MWPC gas
pressure within a few torr. (We would like to keep the MWPC gas
pressure slightly higher than the nitrogen pressure so that the exit
window will slightly bow out to keep it from touching the cathode wire
plane.)

\begin{figure}
\centering
\includegraphics[width=10cm,angle=90]{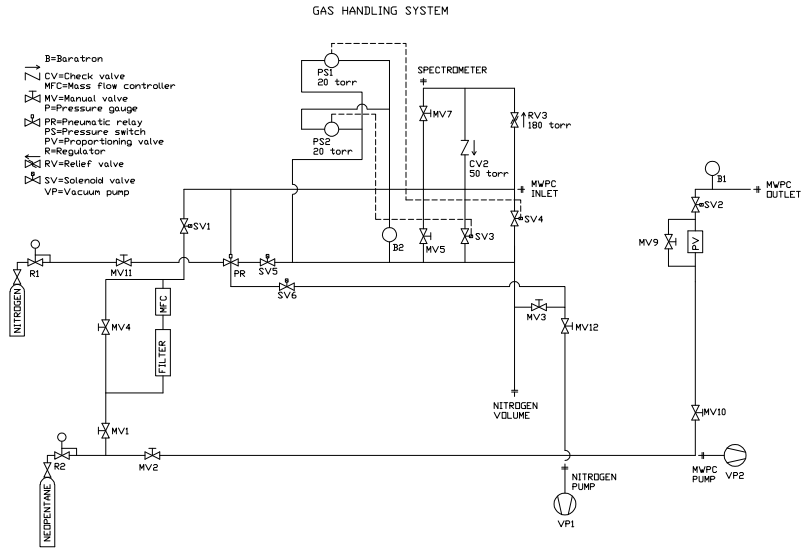}
\caption{Diagram of the actual gas handling system}
\label{fig:gaspanel2} 
\end{figure}

Figure~\ref{fig:gaspanel2} shows the actual gas panel diagram,
including valves and pressure switches that allow us to pump out the
MWPC volume safely and protect the MWPC windows in case of accidents.

\section{Detector performance test and results}
Two identical MWPCs were successfully constructed.  An off-line
detector test was performed using an $^{55}$Fe source to study the
MWPC response to mono-energetic x rays. The pulse height of the
5.9~keV signal was studied as a function of the bias voltage. A
130-keV electron gun was then used to test the MWPC for its response
to an electron beam.  Each of these tests are described in detail
below.

\subsection{Gas property study with an $^{55}$Fe source}
\label{sec:gasproperties}
The response of the UCNA MWPC was first studied by irradiating the
MWPC by 5.9~keV x rays (Mn K x rays) from an $^{55}$Fe source. At this
energy, the photon-matter interaction is dominated by the
photoelectric effect. Since the average energy deposition of
$\beta$-decay electrons in the MWPC is several keV, the total energy
peak of the 5.9~keV x rays provides an almost ideal means to study the
performance of the UCNA MWPC.

Because of the technical and administrative difficulties associated
with placing the $^{55}$Fe source in vacuum, the tests were done using
an arrangement as follows. The MWPC front window was replaced with an
aluminum plate with a 25.4-mm diameter hole. A window made of
12-$\mu$m thick Mylar sheet was glued over the hole, and x rays were
introduced through this hole.

For comparison, we also performed the same tests with the MWPC filled
with 1~atm P10 gas (90\% argon, and 10\% ethane), which is commonly
used as an MWPC gas.

\begin{figure}
\centering
\includegraphics[width=10cm]{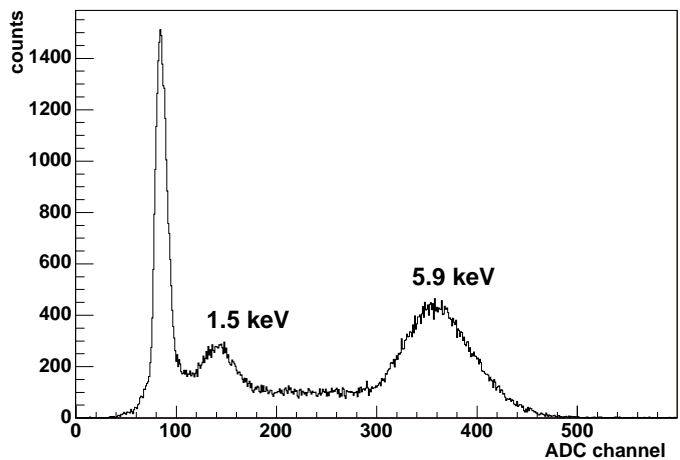}
\caption{Typical x-ray spectrum with the MWPC filled with 100~torr
  neopentane gas. In addition to the Mn K x rays at 5.9~keV, we also
  see the Al K x rays at 1.5~keV, which are generated by Mn K x rays
  incident on the chamber body which is made of aluminum.
} \label{fig:xrayspectrum}
\end{figure}

Shown in Fig.~\ref{fig:xrayspectrum} is a typical x-ray spectrum with
the MWPC filled with 100~torr neopentane gas. In addition to the Mn K
x rays at 5.9~keV, we also see the Al K x rays at 1.5~keV, which are
generated by Mn K x rays incident on the chamber body which is made of
aluminum.

\begin{figure}
\centering
\includegraphics[width=10cm]{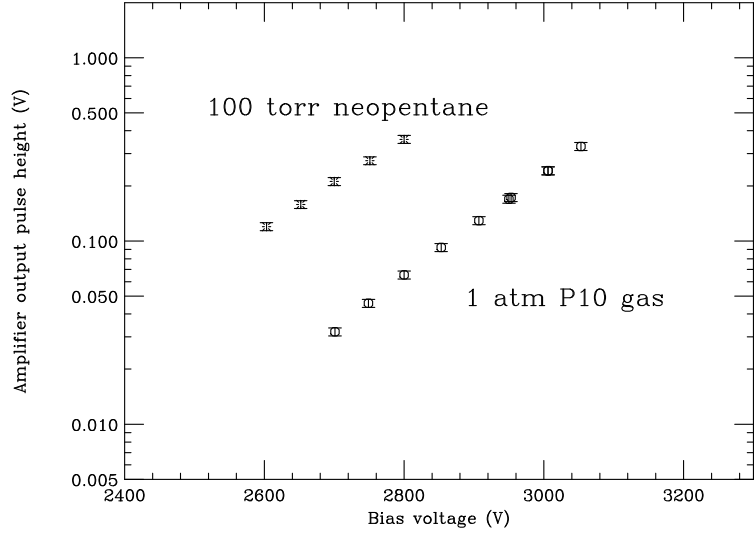}
\caption{Measured pulse height of the amplifier output of the anode
signal for 5.9~keV x rays as a function of the bias voltage with the
MWPC filled with 100 neopentane. Also shown for comparison is the
measured pulse height with the MWPC filled with 1~atm P10 gas.}
\label{fig:pulseheight}
\end{figure}

We also measured the pulse height of the amplifier output of the anode
signal for 5.9~keV x rays as a function of the bias voltage between
the anode and the cathode with the MWPC filled with 100 torr
neopentane. For this measurement, the cathode and the entrance and
exit windows were kept at the same potential for simple
operation. Also, the same measurement was performed with the MWPC
filled with 1~atm P10 gas for a
comparison. Figure~\ref{fig:pulseheight} shows the measured pulse
height as a function of the bias voltage. As seen from the figure, a
reasonable pulse height was obtained with 100~torr neopentane. From a
detection efficiency measurement, the nominal operational voltage was
chosen to be 2700~V.

In principle, the measured pulse height can be related to the gas
multiplication factor as follows.
\begin{equation}
\label{eq:gasgain}
V = G_{\rm amp}E_{\rm xray}\frac{1}{W}eM\epsilon_{\rm cc},
\end{equation}
where $V$ is the size of the amplifier output voltage pulse, $G_{\rm
amp}$ is the gain of the amplifier which converts a charge pulse to a
voltage pulse, $E_{\rm xray}$ is the energy of an x-ray photon, $W$ is
the amount of energy necessary to create one electron-ion pair in the
gas, $e$ is the absolute value of the charge of the electron
($e=1.6\times 10^{-19}$~C), and $M$ is the gas multiplication
factor. $\epsilon_{\rm cc}$ accounts for the reduction in the pulse
height due to the fact that the shaping time of the amplifier is
shorter than the time it takes for the pulse to reach its maximum
amplitude. For this particular geometry, it is estimated that
$\epsilon_{\rm cc}\sim 0.4$. With $E_{\rm xray}=5.9$~keV and $G_{\rm
amp} = 2$~V/pC, and assuming $W=26$~eV for neopentane,\footnote{The
$W$-value for methane (CH$_4$) is known: $W=27.1$~eV. The $W$-value is
typically $20-30$~eV for most gases.} we obtain for the gas
multiplication factor $M\sim 7\times 10^3$ at 2700 V.

\subsection{Tests with a low energy electron gun}
\begin{figure}
\centering
\includegraphics[width=14cm]{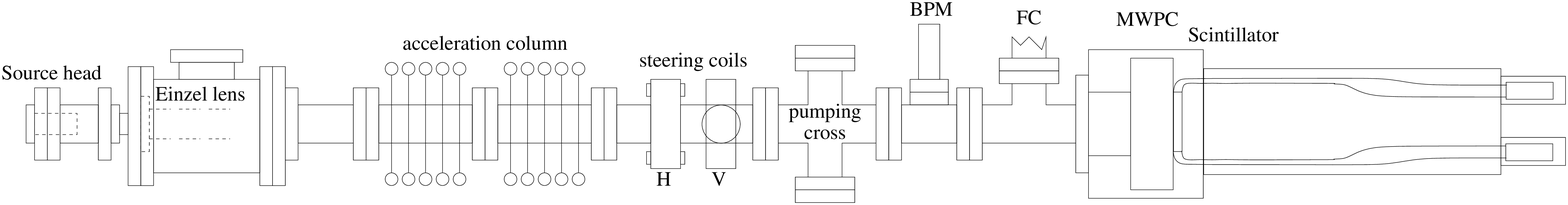}
\caption{Schematic of the setup used to test the MWPC with a 130~keV
  electron beam.}
\label{fig:egunsetup}
\end{figure}

The response of the UCNA MWPC was then studied using the 130~keV
Kellogg electron gun. A schematic of the set up for this test is shown
in Fig.~\ref{fig:egunsetup}. The Kellogg electron gun was able to
provide an electron beam of up to 130~keV and the current could be
varied from a few electrons per second to several microamperes. A more
detailed description of the Kellogg electron gun can be found in
Ref.~\cite{EGUN}.

\begin{figure}
\centering
\includegraphics[width=13cm]{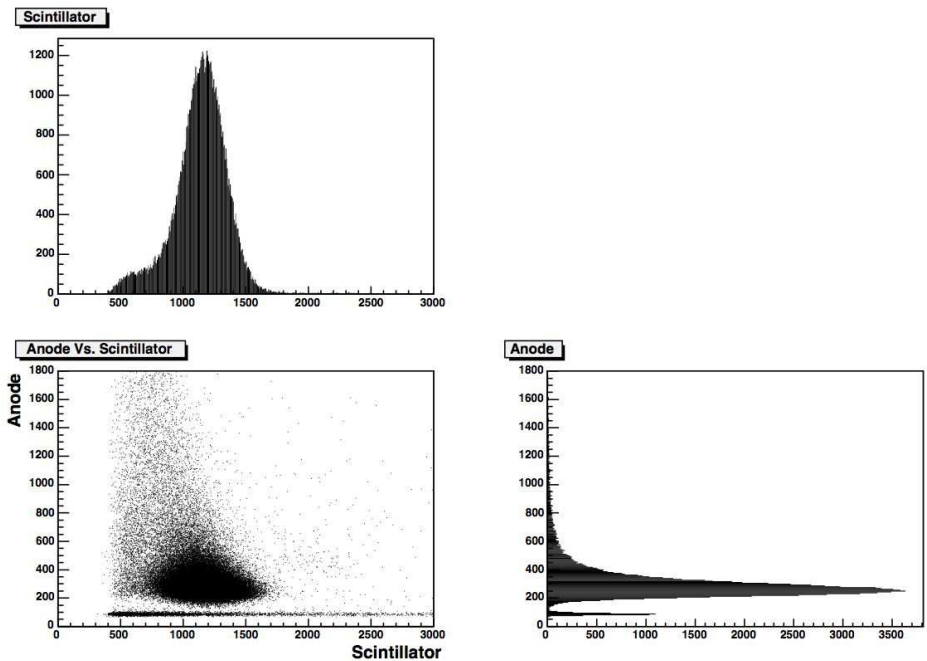}
\caption{Distribution of the scintillator and the MWPC anode pulse
  height with a 130-keV electron beam incident on the detector. The
  trigger is provided by the scintillator. The events where the MWPC
  recoded no signal are due to room background.  }
  \label{fig:egunspectrum}
\end{figure}

For this test, the electron current from the Kellogg electron gun was
adjusted so that the event rate on the MWPC was a few kHz. The trigger
was provided by the plastic scintillator located downstream of the
MWPC. In Fig.~\ref{fig:egunspectrum}, we show pulse height spectra of
the anode signal and the scintillator signal.

\begin{figure}
\centering
\includegraphics[width=10cm]{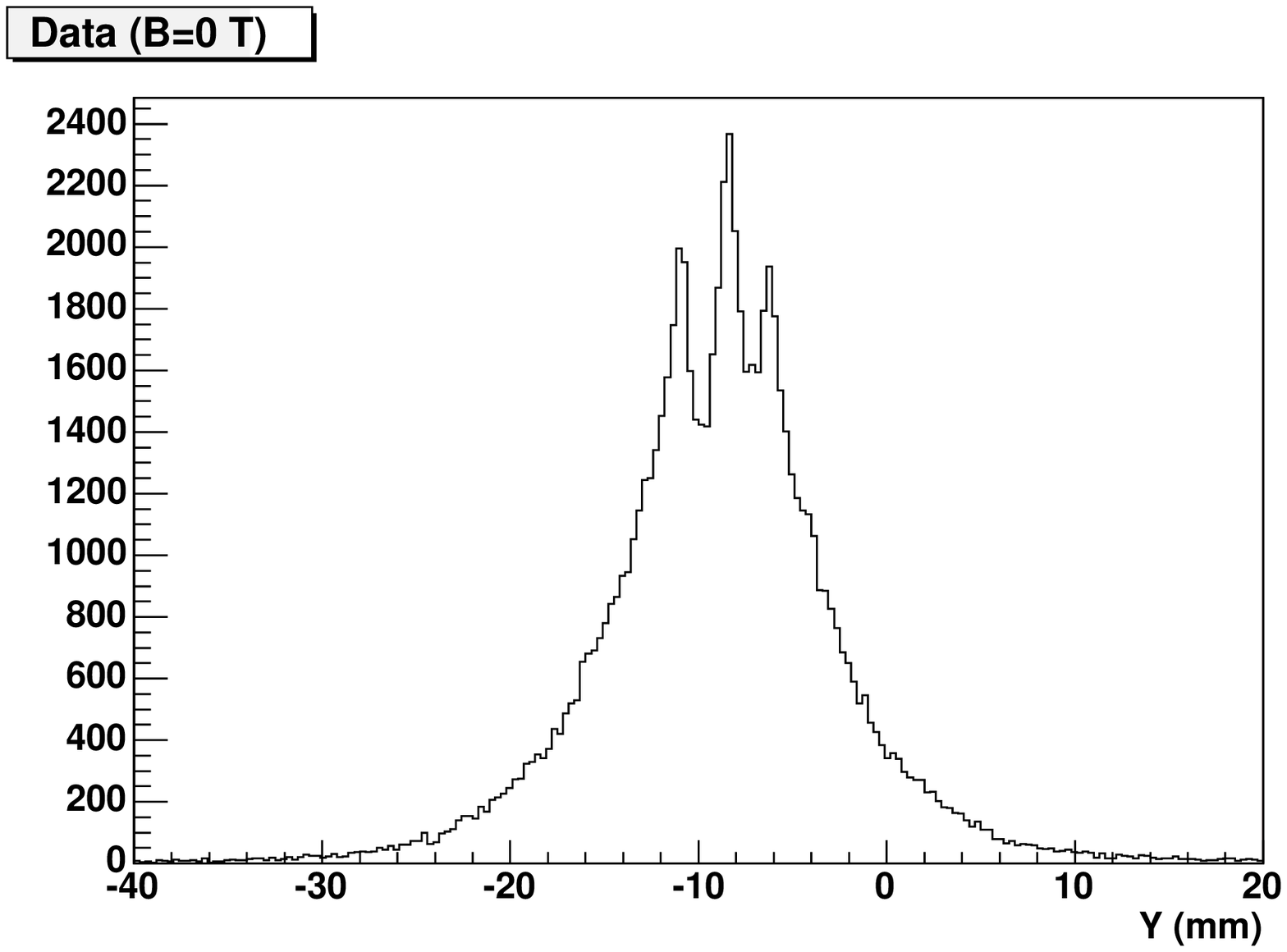}
\includegraphics[width=10cm]{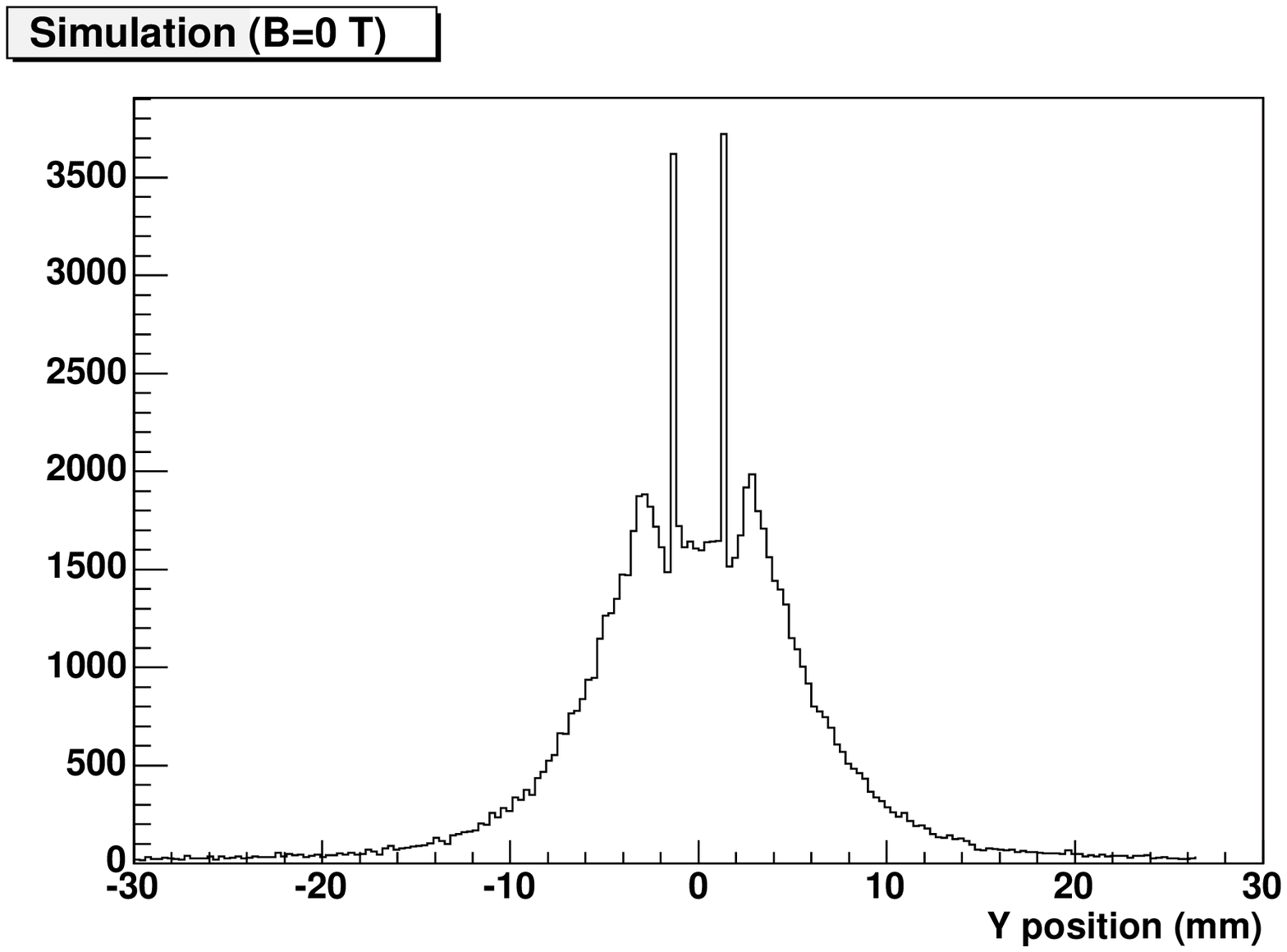}
\includegraphics[width=10cm]{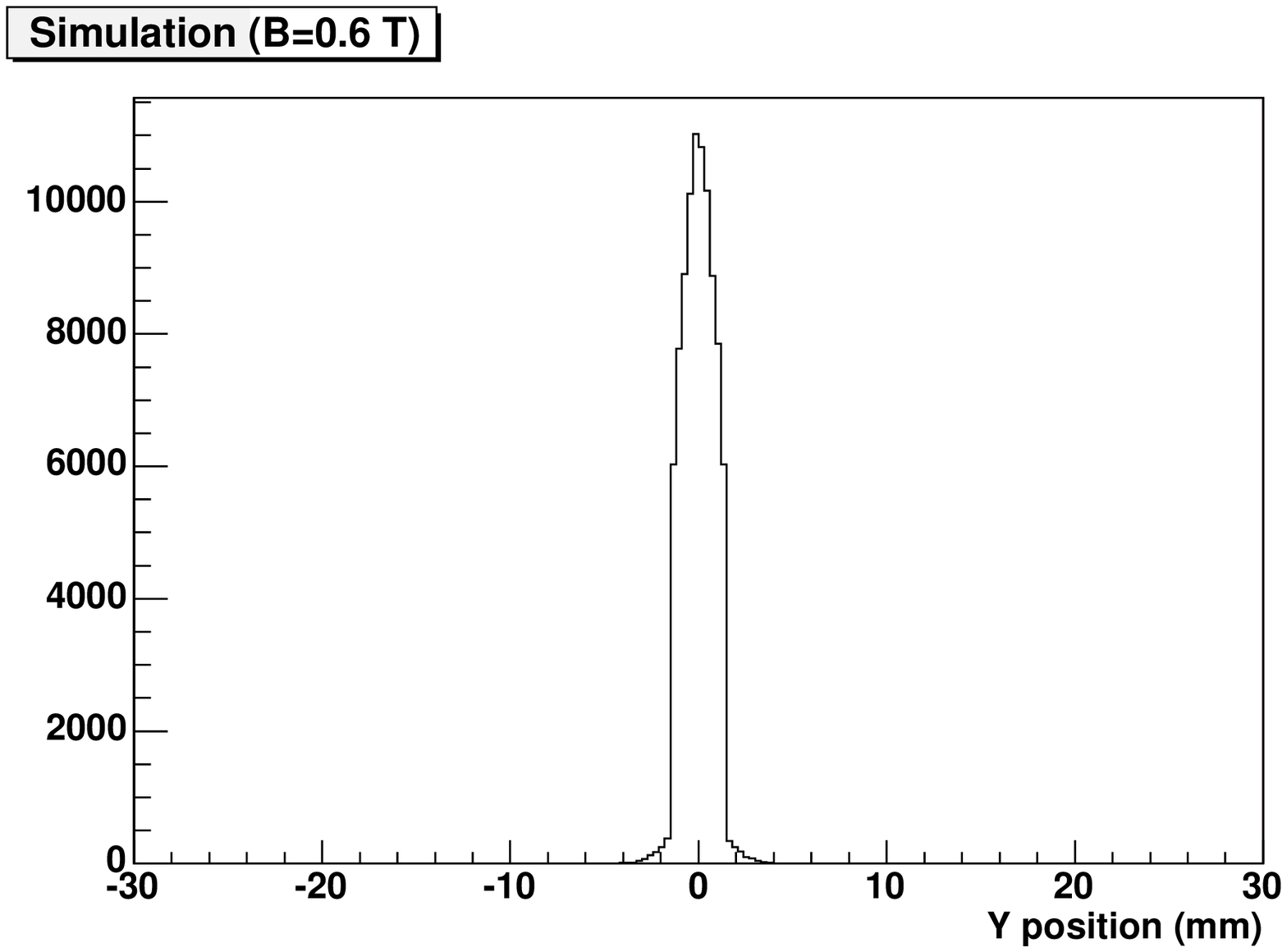}
\caption{Distribution of the reconstructed position of the electron
  hit along the axis perpendicular to the anode wire direction. The
  position was reconstructed from the cathode information. Top panel:
  data taken with an electron beam from the Kellogg electron gun with
  no magnetic field. Middle panel: Monte Carlo simulation with no
  magnetic field. (Note that the effect of electronic noise is not
  included in the simulation and also that the location of the beam
  center with respect to the center of the MWPC is different between
  the simulation and the data.) The bottom panel: Monte Carlo
  simulation with a magnetic field of 0.6 T.  } \label{fig:yposition}
\end{figure}

As discussed in Section~\ref{sec:readout}, the hit position for each
event was reconstructed by first obtaining the center of gravity of
the signals on each cathode plane, and then combining the information
from both cathode planes. The top panel of Fig.~\ref{fig:yposition}
shows the reconstructed position along the axis perpendicular to the
anode wire direction.

Since there is no strong magnetic field along the direction of the
beam (as in the full experiment), the overall position distribution
(full width$\sim 15$~mm) is dominated by multiple scattering in the
front window and the MWPC gas. A Monte Carlo simulation was performed
and it was confirmed that the observed position distribution is
consistent with the simulation (see the middle panel of
Fig.~\ref{fig:yposition}). In addition to this 15~mm wide
distribution, there are narrow spikes. These are due to events in
which only one cell\footnote{An MWPC is divided into cells, each
containing one anode wire at its center. Electric field lines going
through any part of a given cell terminate at the same anode wire.}
received energy deposition. The position of the spikes corresponds to
the location of the anode wires around which the avalanche
occurred. The continuum below the spikes is due to events where more
than one cell received energy deposition. The width of the spikes is
$\sim 0.8$~mm and is the position resolution of the MWPC in the
absence of multiple scattering. This resolution is dominated by
electronic noise. This, combined with simulations, indicates that the
position of the center of the spiral of each beta particle entering
the MWPC can be determined to better than 1.5~mm once the MWPC is put
in the 0.6~T magnetic field of the field expansion region of the SCS
magnet, which will greatly suppress the multiple scattering
effect (see the bottom panel of Fig.~\ref{fig:yposition}). 

\section{Summary}
A new multiwire proportional chamber was designed and constructed for
precision studies of neutron $\beta$ decay angular correlations. Its
design has several novel features, including the use of neopentane as
the MWPC gas and the Kevlar supported thin entrance window. Two such
MWPCs were successfully constructed. Various off-line tests confirmed
that these MWPCs have sufficient sensitivity to low energy $\beta$
particles and allow hit position determination with sufficient
position resolution.

\section{Acknowledgments}
We gratefully acknowledge the technical support of R.~Cortez and
J.~Pendlay. We thank J.~Mohnke for his suggestion of using neopentane
as the MWPC gas, P.~Chan for his help with setting up the wire winding
machine, and E.~Lin and C.~Kessens for their help with the thin window
development. We thank D.~Snowden-Ifft of Occidental College for
generously loaning us his wire winding machine. We thank C.~Morris for
various valuable suggestions. This work was supported by the National
Science Foundation.


\begin{thebibliography}{00}


\bibitem{ABE04}
H. Abele, {\it et al.}, Eur. Phys. J. C {\bf 33}, 1 (2004).

\bibitem{MOR02}
C.~L.~Morris, {\it et al.}, Phys. Rev. Lett. {\bf 89}, 272501 (2002).

\bibitem{SAU04}
A.~Saunders, {\it et al.}, Phys. Lett. {\bf B593}, 55 (2004)

\bibitem{UCNA} The UCNA Collaboration, {\it A proposal for an accurate
  measurement of the neutron spin$-$electron angular correlation in
  polarized neutron beta-decay with ultracold neutrons}, 2000.

\bibitem{PLA06} 
B.~Plaster, {\it et al.}, in preparation.

\bibitem{TAB71}
T.~Tabata, R.~Ito, and S.~Okabe, Nucl. Instr. and Meth. {\bf 94}, 509 (1971). 

\bibitem{LAZ98}
D.~Lazic, {\it et al.}, Nucl. Instr. and Meth. A {\bf 410}, 159 (1998).

\bibitem{BRE79}
A.~Breskin, R.~Chechik, and N.~Zwang, Nucl. Instr. and Meth. {\bf 165},
125 (1979).

\bibitem{MYL}
Mylar$^{\textregistered}$ is a registered trademark of DuPont Teijin Films.

\bibitem{MCS} Multi Channel Systems, http://www.multichannelsystems.com

\bibitem{KUZ87}
M.~Kuze, {\it et al.} Jpn. J. Appl. Phys. {\bf 26}, 1348 (1987).

\bibitem{SNO00}
D.~P.~Snowden-Ifft, C.~J.~Martoff, and J.~M.~Burwell, Phys. Rev. D {\bf
61}, 101301(R) (2003).

\bibitem{Ogura}
PHT Permanent Magnet Clutch/Brake Model PHT-0.05s, Ogura Industrial
Corp., {\tt http://www.ogura-clutch.com/}.

\bibitem{SAU77}
F.~Sauli, CERN Report 77-09 (1977).

\bibitem{ARA} 
Araldite$^{\textregistered}$ is a registered trademark
of Huntsman Advanced Materials.

\bibitem{DIE} Dielectric Sciences Inc., 88 Turnpike Rd., Chelmsford,
  MA 01824-3526.
 
\bibitem{CAEN} CAEN S.p.A. Construzioni Apparecchiature Ellecttroniche
Nucleari, http://www.caen.it/.


\bibitem{KEV}
Kevlar$^{\textregistered}$ is a registered trademark of DuPont.

\bibitem{ATE93}
L.~G.~Atencio, C.~L.~Morris, and C.~P.~Sadler, Nucl. Instr. and
Meth. A {\bf 334}, 643 (1993).

\bibitem{MKS} 
MKS Instruments, Inc., http://www.mkinst.com/.

\bibitem{BAR}
Baratron$^{\textregistered}$ is a registered trademark of MKS
Instruments, Inc.

\bibitem{EGUN}
J.~W.~Martin {\it et al.}, Phys. Rev. C {\bf 68},
055503 (2203).

\end{thebibliography}
\end{document}